# Asymmetric Position-Based Quantum Cryptography

Muhammad Nadeem
*muhammad.nadeem@seecs.edu.pk*
*School of Electrical Engineering and Computer Science*
*National University of Sciences and Technology*
*H-12 Islamabad, 44000, Pakistan*

A large number of quantum location verification protocols have been proposed. All existing protocols in this field are based on symmetric cryptography where verifiers and the prover use the same secret key. The prover obtains secret key from different verifiers, in pieces through public channels, which may result in security compromise. In this paper, we give a formulism to prove information-theoretic position-based quantum cryptography is possible in asymmetric quantum setting. We propose a quantum location verification protocol based on asymmetric quantum cryptographic where two different keys will be used.

## I. INTRODUCTION

Quantum cryptography offers unconditional security in communication [1,2, 3] through quantum key distribution [4,5]. Recently, different authors [6-14] claimed same information-theoretic security in position-based quantum cryptography (PBQC); identification and position verification of a prover by sending and receiving quantum/classical signals from various verifiers at distant reference stations. The first quantum scheme for position verification was proposed by Kent, Munro, Spiller and Beausoleil in 2002 under the name of "quantum tagging" and later a US patent was granted in 2006 [6]. The central task of position-based cryptography introduced by Chandran et all [15] is position verification; a prover proves to a set of verifiers located at certain distant reference stations that he/she is indeed at a specific position. They proved that unconditional security in classical PBC is impossible because of cloning. The eavesdroppers can copy classical information, manipulate and send response to the verifiers before a honest prover.

Buhman *et all* [14] claimed that if honest prover neither share any secret information with verifiers nor he has any advantage over eavesdroppers beyond his position in the environment fully controlled by eavesdroppers with arbitrarily large pre-shared entanglement, then no information-theoretic position-based quantum scheme is possible at all. They showed that security of any position-based quantum cryptographic protocol can be destroyed by eavesdroppers through teleporting quantum states back and forth and performing instantaneous nonlocal quantum computation, an idea introduced by Vaidman [16]. However, they proved that if eavesdroppers do not share any entanglement (NO-PE model), then secure PBQC is possible. Furthermore, S. Beigi and R. Konig showed that if eavesdroppers posses an exponential (in n) amount of entanglement then they can successfully attack any PBQC scheme where verifiers share secret n-bit string [17].

Some authors also proposed position-based quantum cryptographic protocols where prover and verifiers posses some data secret from adversary, instead of communicating all information through public channels. Kent proposed that secure PBQC is possible if prover and verifiers pre-share some classical bit string unknown to eavesdroppers [11]. Moreover, as proposed in our previous work [12], position-based quantum cryptography can be made secure by sharing entangled data and using entanglement swapping [18] instead of sending information through public channels. Buhrman *et all* [14] proposed a protocol, $PV_{BB84}^{\varepsilon}$ EPR version, where one of the verifiers share an entangled state with the prover. The protocol also requires a secret bit string shared between the verifiers who send this secret information to the prover publically. Moreover, it is worth mentioning that R. A. Malaney [13], proposed a large class of position-verification protocols where different distant verifiers and the prover share entangled data. His work was granted US patent in 2012.

Currently, all existing position-based quantum cryptographic protocols are based on symmetric setting; verifiers send messages encrypted by a secret key and the prover respond to them by decrypting cipher texts with the same secret key. The prover obtains secret key from different verifiers, in pieces through public channels, which may result in security compromise.

In this work we propose a quantum location verification protocol in different setting, asymmetric quantum cryptography. In asymmetric quantum setting, verifiers and the prover use different keys. Any secret information sent on public channels will be securely encrypted by quantum public-key while the encrypted information will be decrypted by the private key. Proposed protocol is unconditionally secure in general and is equipped to handle entanglement base attacks in particular. Our paper is organized as follows. In section II, we discuss asymmetric quantum cryptography while in section III, we will introduce our protocol. Finally, we summarize the paper in section IV.

## II. ASYMMETRIC QUANTUM CRYPTOGRAPHY

Asymmetric quantum cryptography is based on quantum one-way functions (OWF). F(x) is a one-way function if the map x→F(x) is easy to compute but impossible to invert; F(x)→x without knowing x. The idea of quantum OWF was first introduced in [19, 20], where authors proposed quantum finger printing and quantum digital signatures. They showed that quantum OWF can be obtained by mapping all classical bit strings S of length L to quantum state $|\psi_S\rangle$ of n qubits. Later, G. M. Nikolopoulos presented an asymmetric quantum cryptographic scheme by mapping integer number s to single qubit state $|\psi_s\rangle$ [21]. They showed the map $s \to |\psi_s\rangle$ acts as a quantum OWF and carry trapdoor information; only authorized users can invert mapping s→$|\psi_s\rangle$ but it remains impossible for others. It can be achieved via following steps:

*1). One way function:* Let $\{|0_z\rangle, |1_z\rangle\}$ is basis set on the x-z plane of Bloch sphere and an authorized user prepares a T-qubit state $|0_z\rangle^{\otimes T}$. Each integer $s_i$, from secret classical string $S = \{s_1, s_2, \ldots s_T\}$ of length T, can be mapped with corresponding rotated (about y-axis) qubit in the state

$$|\psi_S(\theta_t)\rangle = \otimes_{i=1}^{T} R^{(i)}(s_i\theta_t)|0_z\rangle^{\otimes T} \qquad (1)$$

That is,

$$s_i \to |\psi_{s_i}(\theta_t)\rangle = \cos(s_i\theta_t)|0_z\rangle + \cos(s_i\theta_t)|1_z\rangle \qquad (2)$$

Where $R^{(i)}(s_i\theta_t)$ is the rotation operator, *t* is any arbitrary secret integer and $\theta_t = \pi/2^t$. If *t* >> 1 (or $\theta$ << 1), number of non-orthogonal states increases and it becomes impossible to differentiate them: distance between nearest neighbors $\sqrt{1 - |\langle\psi_s(\theta_t)|\psi_{s+1}(\theta_t)\rangle|^2}$ approaches to zero. Moreover, only one bit of classical information can be obtained from single qubit [22] while T bits are required to identify any randomly chosen $s_i$ from T-bit string S, Hence, the map $s \to |\psi_s\rangle$ acts as a quantum OWF provided *t* >> 1.

Asymmetric Position-Based Quantum Cryptography*2). Trapdoor information:* Suppose two rotations, first $R(s\theta_t)$ and then $R(m\theta_t)$, are applied on the same qubit where *s* and *m* are random integers such that $s + m = n$ mod $2^t$. If $t >> 1$, only authorized user knowing secret integer *s* and *t* can find *m*. Even if adversaries gain substantial amount of information about *n*, it is not possible to deduce m without knowing *s* and *t*. Hence the map $s \rightarrow |\psi_s\rangle$ acts both as quantum one-way and trapdoor function.

We will follow the same quantum scheme [21] in next section and show that position-based cryptography can be made unconditionally secure in an asymmetric setting.

### III. ASYMMETRIC QUANTUM LOCATION VERIFICATION

In this work, we assume that the location of the honest prover and reference stations are secure from adversary; enabling them to store and hide the quantum data and process. We also assume that the reference stations are trusted and known to each other. However, quantum/classical channels are not secure; neither between the prover and verifiers nor between different verifiers. Moreover, there is no bound on storage, computing, receiving and transmitting powers of eavesdroppers. In short, eavesdroppers have full control of environment except prover's location and reference stations. We also assume that all reference stations and the prover has fixed position in Minkowski space-time where all verifiers have précised and synchronized clocks. Finally, we suppose that signals can be sent between prover and reference stations at the speed of light. While the time for information processing at location of the honest prover and reference stations is negligible.

For simplicity, we assume that the honest prover P is at a distance d from all reference stations $R_1, R_2,\ldots\ldots R_N$. Explicit procedure of our protocol follows:

1). The honest prover P prepares a T-qubit state $|0_z\rangle^{\otimes T}$ and a pair of public-private keys. His private key is classical, d = (*t*, *S*), where *t* >>1 is positive integer and $S = \{s_1, s_2, \ldots s_T\}$ is string of length T. While his public key is T-qubit quantum state $|\psi_S(\theta_t)\rangle$, such that

$$|\psi_S(\theta_t)\rangle = \otimes_{i=1}^{T} R^{(i)}(s_i\theta_t)|0_z\rangle^{\otimes T} \quad (3)$$

where $\theta_t = \pi/2^t$ and $s_i$ represents the $i^{th}$ bit of string S.

2). P produces N copies of his public key and sends to N verifiers at distant reference stations $R_1, R_2,\ldots\ldots R_N$. As quantum sate $|\psi_S(\theta_t)\rangle$ is known to P, he can produce multiple copies of his public key without violating no-cloning theorem.

3). Every verifier at station $R_i$ encrypt an $r_k$-bit message $m_k = \{m_1, m_2, \ldots m_{r_k}\}$ with prover's public key without changing the order of its qubits. Here $m_i \in [0,1]$ and $r_k$ < T. At time t = 0, all verifiers send the encrypted message to prover P simultaneously:

$$|\psi_{S,m_k}(\theta_t)\rangle = \otimes_{i=1}^{T} R^{(i)}(m_i\pi)R^{(i)}(s_i\theta_t)|0_z\rangle^{\otimes T} \quad (4)$$

4). P receives messages from all the verifiers at the same time. He decrypts the messages with his private key by applying inverse rotations $R^{(i)}(s_i\theta_t)^{-1}$ and gets

$$|\psi_{S,m_k}(\theta_t)\rangle = \otimes_{i=1}^{T} R^{(i)}(m_i\pi)|0_z\rangle^{\otimes T} \quad (5)$$

He measures each qubit of encrypted message in $\{|0_z\rangle, |1_z\rangle\}$ basis and sends the messages to corresponding verifiers at stations $R_1, R_2, \ldots\ldots R_N$.

5). If all the verifiers agree, identity of P is validated from the announced results. Besides, the position of P can be verified by checking the time elapsed for response from P, t = 2d/c, after sending encrypted message.

## IV. CONCLUSION

In this paper, we gave a formulism to prove information-theoretic position-based quantum cryptography is possible in asymmetric quantum setting. Security of our protocol relies on the fact that no secret information, which could help in decrypting cipher text, is sent directly through public channels but is encrypted by quantum public-key. Cipher text can only be decrypted by authorized prover having private key. In short, our asymmetric position verification protocol remains secure under known entanglement base attacks even if eavesdroppers have infinite amount of pre-shared entanglement and power of non-local quantum measurements in negligible time. Moreover, as far as security of the used public-private keys and asymmetric quantum scheme is concerned, detailed security analysis can be found in [21, 22].

It is worth mentioning that in asymmetric PBQC setting, quantum/classical channels between distant verifiers need not to be secure. While, all previously proposed symmetric PBQC schemes rely on pre-supposed secure channels between distant verifiers. It reflects that asymmetric settings in PBQC have advantages over symmetric one. We hope this paper will give motivations for the search of more practical and robust asymmetric position-based quantum schemes.